\documentclass[conference]{IEEEtran}
 \usepackage{multirow}
\usepackage{adjustbox}
\usepackage{amsmath}
\usepackage[HTML]{xcolor}
\definecolor{myred}{HTML}{C00000}
\usepackage[colorlinks=true, linkcolor=myred, citecolor=blue, urlcolor=black]{hyperref}
\usepackage{cleveref}
\crefformat{section}{#2§#1#3}
\crefname{section}{§}{§§}
\usepackage{booktabs}
\usepackage{amssymb}
\usepackage{textgreek}

\ifCLASSINFOpdf
\else
\fi
\begin{document}
%

\title{Temporal Risk on Satellites}


\author{\IEEEauthorblockN{Shiqi Liu}
	\IEEEauthorblockA{George Mason University\\
		sliu38@gmu.edu}
	\and
	\IEEEauthorblockN{Kun Sun}
	\IEEEauthorblockA{George Mason University\\
		ksun3@gmu.edu}}
	

%


\IEEEoverridecommandlockouts
\makeatletter\def\@IEEEpubidpullup{6.5\baselineskip}\makeatother
\IEEEpubid{\parbox{\columnwidth}{
		{\fontsize{7.5}{7.5}\selectfont Workshop on Security of Space and Satellite Systems (SpaceSec) 2026 \\
			23 February 2026, San Diego, CA, USA \\
			ISBN 978-1-970672-02-2 \\
	\url{https://dx.doi.org/10.14722/spacesec.2026.23050}  \\
			\url{www.ndss-symposium.org}}
}
\hspace{\columnsep}\makebox[\columnwidth]{}}

\maketitle

\begin{abstract}
Satellite vulnerabilities change over time as orbits shift, power margins tighten, and the space environment deteriorates. However, most cybersecurity risk frameworks still treat threats as static. In practice, the same exploit can be far more damaging during a critical maneuver than during routine operations. We propose a temporal risk assessment framework that makes time an explicit axis in satellite security analysis. It extends existing adversary behavior taxonomies with a five-dimensional temporal capability model and estimates exploitation difficulty across distinct temporal windows of a mission. Rather than producing a single risk score, the framework outputs a series of time-indexed likelihood–impact matrices. It discretizes missions into operationally meaningful time windows and environmental bands to show when systems are most exposed. This view helps operators avoid scheduling sensitive operations in high-risk periods and align defensive resources with a threat landscape that shifts over time.
\end{abstract}


%
\IEEEpeerreviewmaketitle

\section{Introduction}

Satellite systems~\cite{starlink2025official} now underpin global communications, navigation, and Earth observation, making them critical infrastructure. As these systems grow more sophisticated and more interconnected, their attack surface expands. Their predictable orbits, limited physical protection, and complex supply chains create persistent opportunities for adversaries~\cite{vanlyssel2025spychain}. The growing use of Commercial Off-The-Shelf (COTS) hardware and inexpensive ground station services further lowers barriers to entry. These trends make it easier and cheaper to interact with and attack satellite systems~\cite{remy2025sok}. Recent incidents, such as the 2022 KA-SAT cyberattack~\cite{viasat2022attack} that disrupted communications across Europe, show that satellite security threats are not hypothetical.

Prior work has emphasized spatial and architectural aspects, but the \emph{timing} of attacks can matter as much as \emph{what} attackers do and \emph{how} they do it. Satellite operations are governed by time-dependent phenomena. Orbital mechanics determines when ground stations can communicate with spacecraft. Mission-critical operations such as orbital maneuvers, payload activations, and data downlinks run within narrow time windows. Adversaries who understand these temporal patterns can time operations to increase impact. Recent studies~\cite{willbold2025space} show that space-specific effects, such as radiation-induced bit flips, can interfere with defenses such as Control Flow Integrity (CFI). During South Atlantic Anomaly (SAA) passages or geomagnetic storms, elevated single-event upset (SEU) rates increase the residual risk of mitigation failure. Complementing these radiation-induced effects, a recent study on energy-drain attacks in satellite Internet constellations also supports this view. Low Earth orbit (LEO) satellites alternate between sunlight and eclipse and rely on batteries during eclipse. The depth of battery discharge has a major effect on lifetime. The STARMELT attack~\cite{zhang2023energy} sends crafted traffic through a victim satellite to keep its communication subsystem in a high-power state instead of allowing it to sleep. It exploits predictable orbits and access patterns to align traffic with eclipse periods and to maximize battery drain.

Existing space cybersecurity risk frameworks, such as SPARTA~\cite{ear2024towards}, are widely used for threat modeling and for mapping techniques to NIST SP~800-53 controls~\cite{kurii2022analysis}. However, current approaches treat risk as essentially \emph{time-invariant}. Each technique receives a single risk score that implicitly assumes its feasibility and impact remain constant throughout the mission. The absence of an explicit temporal dimension in risk assessment creates three gaps: \emph{(i) Incomplete likelihood estimation}. Existing models do not capture adversaries’ temporal capabilities or how the difficulty of executing an attack varies across time windows. \emph{(ii) Impact underestimation}. Static risk scores cannot reflect how system vulnerability changes with operational context. \emph{(iii) Misaligned defenses}. Without temporal risk profiles, defenders cannot prioritize protection during high-risk windows or schedule sensitive operations outside predictable vulnerability periods.

We present a \emph{temporal\footnotemark~risk framework} for satellite cybersecurity that treats timing as a primary concern in threat assessment. Building on SPARTA's Notional Risk Scores (NRS), we extend the standard 5$\times$5 matrix over likelihood and impact into a family of matrices indexed by time. Each matrix evaluates risk for specific mission time windows and environmental bands (e.g., eclipse vs.\ non-eclipse, geomagnetic storm severity levels). The framework fits into existing risk management workflows. Analysts can add temporal considerations on top of established ATT\&CK-based assessments. By making explicit \emph{when} attacks are feasible, high-impact, or difficult to attribute, the framework gives operators a time-aware view of risk that current practices lack.

\footnotetext{In this paper, we use the term \emph{temporal} to refer to the dependency on physical time, orbital mechanics, and mission schedules. This is distinct from the \emph{Temporal Metrics} in CVSS, which refer to the lifecycle status of a vulnerability (e.g., exploit maturity or patch availability).}

\section{Background \& Related Work}

\subsection{Attacks Against Satellite Systems}

Satellite attack surfaces span the ground, link, user, and space segments~\cite{remy2025sok,peled2023evaluating}. Adversaries often gain a foothold in the ground segment through ground networks, Ground Station as a Service (GSaaS) platforms, or operational tooling~\cite{willbold2023space,viasat2022attack}. The compromise can then cascade to space assets. Effective defense therefore requires a holistic analysis of end-to-end mission control and data flows instead of treating the spacecraft in isolation.

Within the link and user segments, over-the-air jamming and spoofing~\cite{tibaldo2025gnss,salkield2023satellite,oligeri2020gnss} remain prominent, and the modem and terminal layers are frequent weak points. Teardown and measurement studies that focus on low Earth orbit terminals and user equipment~\cite{wouters2022glitched,willbold2024vsaster,smailes2023dishing} have revealed design and implementation defects. They also show concrete paths to exploitation.

On the ground segment, falling ground station costs and the emergence of GSaaS~\cite{aws2025groundstation,azure2025orbital} have eroded physical and economic barriers. It is now easier for attackers to interact with assets in orbit and to trigger software vulnerabilities. Empirical analyses of firmware from satellites in orbit~\cite{scharnowski2023case,willbold2023space} show that some systems lack protection of telecommand channels and lack access control. They also exhibit multiple software defects. These weaknesses show that relying on obscurity alone is fragile.

Isolation and recovery capabilities in orbit are therefore central to resilience. Small satellites increasingly run full operating system stacks, for example Linux-based payload computers. Recent work~\cite{marra2024feasibility,yadav2024orbital} has begun to evaluate application sandboxes as a practical isolation layer for CubeSat-class missions and to document feasibility and engineering challenges for integration in orbit. In parallel, proposals for independent health monitoring and out-of-band telemetry channels~\cite{young2010initial,wyatt1997overview} highlight the prevalence of subsystem faults. They also show the importance of post-incident diagnosis and recovery for limiting the blast radius of cyberattacks.

\subsection{MITRE ATT\&CK in Space}

\begin{figure}[!htbp]
  \centering
  \includegraphics[width=1.8 in]{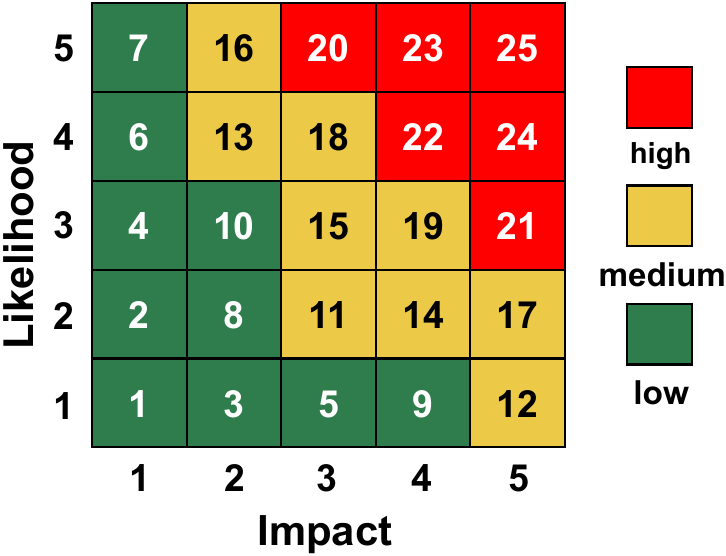}
  \caption{Risk Matrix Representation~\cite{aerospace_sparta_website}.}
  \label{fig:matrix}
\end{figure}

MITRE ATT\&CK~\cite{mitre_attack_website} is a matrix-based knowledge base that provides a common lexicon of adversary tactics, techniques, and procedures (TTPs). It spans the Enterprise, Mobile, and Industrial Control Systems matrices and is widely used for adversary emulation and detection coverage assessment. In recent years, researchers have begun to adapt ATT\&CK-based TTP matrices to the space domain. The Aerospace Corporation SPARTA framework~\cite{ear2024towards} defines a space cyber tactics and techniques matrix. Its Notional Risk Scores (NRS) place each technique in a 5$\times$5 matrix over likelihood and impact, as shown in \autoref{fig:matrix}. Likelihood is estimated from adversary motivation, exploitation difficulty, and capability tier. Impact covers mission disruption, loss of control or availability, safety effects, financial loss, and national security consequences. Cell values from 1 to 25 are grouped into low, medium, and high and drive the selection of mapped countermeasures and NIST SP~800-53 controls~\cite{kurii2022analysis}.
 
\section{Methodology}

Building on the SPARTA Notional Risk Scores summarized above, we now describe how to extend the static 5$\times$5 risk matrix into a temporal form. For each technique, we first examine how likelihood and impact vary across the relevant time windows. We assign 1–5 levels for likelihood and impact to each window.

For likelihood assessment, we apply time-dependent adjustments to \emph{Exploitation Difficulty} and \emph{Adversary Capabilities}. These adjustments reflect temporal accessibility and adversary timing proficiency. \emph{Motivation} retains its original meaning. It is primarily driven by system criticality and the resulting attractiveness of the target. Favorable time windows and exploitation costs may influence how an adversary prioritizes targets. They do not change the intrinsic appeal of the target. It is worth noting that in traditional static frameworks~\cite{ear2024towards}, analysts often implicitly account for narrow time windows by lowering the overall likelihood score. While this produces a valid time-averaged assessment, it can mask periods of acute vulnerability.

Impact is also modulated by time. Time windows influence the realized consequences and severity of a successful attack. This effect depends on the mission and on the specific program. Time windows do not change the impact scoring framework itself. The category scheme (I1--I5) and its semantics, such as mission interruption, partial degradation, or loss of mission, remain unchanged. Time determines which level in that framework a specific attack scenario should map to. For example, the same jamming attack might be assessed as I2 (recoverable data delay) during routine operations but as I5 (battery depletion that terminates the mission) during an eclipse-critical period. Both ratings use the same impact framework. The difference arises from how vulnerable the system is in the respective time windows.

\begin{table*}[t]
  \centering
  \caption{Minimal temporal capability sets $M_T$ for the seven attacker tiers.}
  \label{tab:capability}
  \begin{adjustbox}{max width=\linewidth}
  \begin{tabular}{@{}clcl@{}}
    \toprule
    Tier & Representative adversary type & $M_T$ & Intuitive description \\
    \midrule
    1 & Script kiddies
      & $\{A\}$
      & Uses off-the-shelf tools in rough time windows, with no real prediction, sensing, or precise timing. \\
    2 & Hackers for hire
      & $\{A, F\}$
      & As above, plus basic forecasting of access windows using public orbit and pass-prediction tools. \\
    3 & Small hacker teams
      & $\{A, F, S\}$
      & Add continuous sensing (e.g., spectrum monitoring) to refine when the target is accessible and responsive. \\
    4 & Insider threats
      & $\{A, F, S, Y\}$
      & Combine high-quality internal state data with the ability to align actions to precise operational time slots. \\
    5 & Large well-organized teams
      & $\{A, F, S, Y, H\}$
      & Possess the full temporal capability set, including deliberate timing for statistical and operational hiding. \\
    6 & Highly capable state actors
      & $\{A, F, S, Y, H\}$
      & Same temporal capability set as Tier~5, but applied at greater technical depth and operational scale. \\
    7 & Most capable state actors
      & $\{A, F, S, Y, H\}$
      & As Tier~6, further extended to coordinated, multi-theater, multi-constellation campaigns. \\
    \bottomrule
  \end{tabular}
  \end{adjustbox}
\end{table*}

\subsection{Temporal Adversary Capabilities}

Existing space cybersecurity risk frameworks~\cite{ear2024towards,bailey2021cybersecurity} use a seven-tier attacker model to quantify adversary capability. The tiers range from script kiddies to the most capable state actors. Each tier is described along seven generic dimensions, such as vulnerability discovery and exploitation. To remain comparable to these frameworks, we keep the seven-tier categorization. Instead, we map \emph{time} back into attacker capability itself. We introduce a temporal mastery profile $\langle A, F, S, Y, H\rangle$ (actuation, forecasting, sensing, synchronization, and hiding) as an additional threshold for tier assignment. For each attacker class we define the minimum required set of temporal capabilities, as shown in \autoref{tab:capability}. This preserves the generality of the original tiers while embedding temporal proficiency into attacker capability metrics. It also lets us compare timing-driven adversarial operations across tiers.

\noindent \textbf{Actuation (A)} captures whether an adversary can cause effects on the target within a given access window. It includes RF power shaping and directed jamming at the waveform layer and malicious uplinks at the protocol and command layers. In extreme cases, it also covers physical actions such as laser dazzling. The exploitable window is bounded by the visibility window. This window is set by practical engineering constraints such as line-of-sight (LoS), antenna pointing, and radio frequency (RF) chain availability.

\noindent \textbf{Forecasting (F)} turns observations into time-localized predictions. It applies simplified perturbation models for orbital propagation (SGP4 for near-Earth and SDP4 for deep-space objects) together with link geometry calculations. These models estimate future access windows, eclipse ingress and egress times, and link budget extrema. Forecasting also uses space weather and radiation models to anticipate SAA crossings and geomagnetic storm intervals.

\noindent \textbf{Sensing (S)} provides near-real-time awareness of system state. It includes passive monitoring of unencrypted beacons and telemetry, spectrum and power side-channels, and optical or radar observations. It also covers high-quality, low-latency telemetry from compromised ground assets or from attacker-controlled ground stations. These sources can be fused with third-party orbital data and space weather feeds to produce time-aligned situational awareness.

\noindent \textbf{Synchronization (Y)} is the ability to align actions with a chosen window in time. It depends on a stable time base and a well-characterized end-to-end delay budget. In practice, this includes disciplining clocks via GPS or NTP, modeling and compensating propagation and processing delays, and phase-locking actions to onboard timing or ground-station handovers. Without Synchronization (Y), even strong forecasting and sensing capabilities rarely produce reproducible effects.

\noindent \textbf{Hiding (H)} aims to make induced effects hard to distinguish from noise in statistics. It keeps observable outcomes within expected statistical bounds and mimics natural failure signatures such as SEU-like event rates. It also times actions to coincide with space weather activity, Sun outage periods, or maintenance windows. A high level of Hiding (H) assumes a baseline model of nominal and anomalous behavior and competent statistical modeling.

\subsection{Temporal Exploitation Difficulty}

Following the SPARTA framework~\cite{ear2024towards}, we retain \emph{Exploitation Difficulty} as a component of technique likelihood. We decompose it into static and temporal components. The static component corresponds to the original notion in SPARTA. It captures code complexity, required tooling, and non-temporal preconditions. The temporal component, which we call \emph{Temporal Exploitation Difficulty (TED)}, is orthogonal to this static part. TED measures how hard it is to exploit a technique at the \emph{right time}, assuming the attacker already knows how to execute the technique in principle. We distinguish two sources of temporal difficulty:

\noindent \textbf{Intrinsic properties of the time window.}
These properties come from the system’s orbital visibility and mission operations timeline. They are independent of any specific adversary. Key factors include the duration of each usable window, how often such windows occur, how predictable their timing is, and whether the opportunity is one-shot or repeatable. For example, routine downlink passes offer frequent and predictable windows with considerable slack. In contrast, windows driven by geomagnetic storm activity or contingency operations are less predictable in practice. Intrinsic properties of time windows can often be analyzed with system-level tools. Orbit propagation and link geometry models~\cite{rhodes2019skyfield} capture access and eclipse windows. Radiation and SAA models~\cite{sawyer1976ap} capture regions with enhanced particle flux. Statistical models capture the frequency and duration of rare combined states.

\begin{figure*}[!t]
  \centering
  \includegraphics[width=6.5 in]{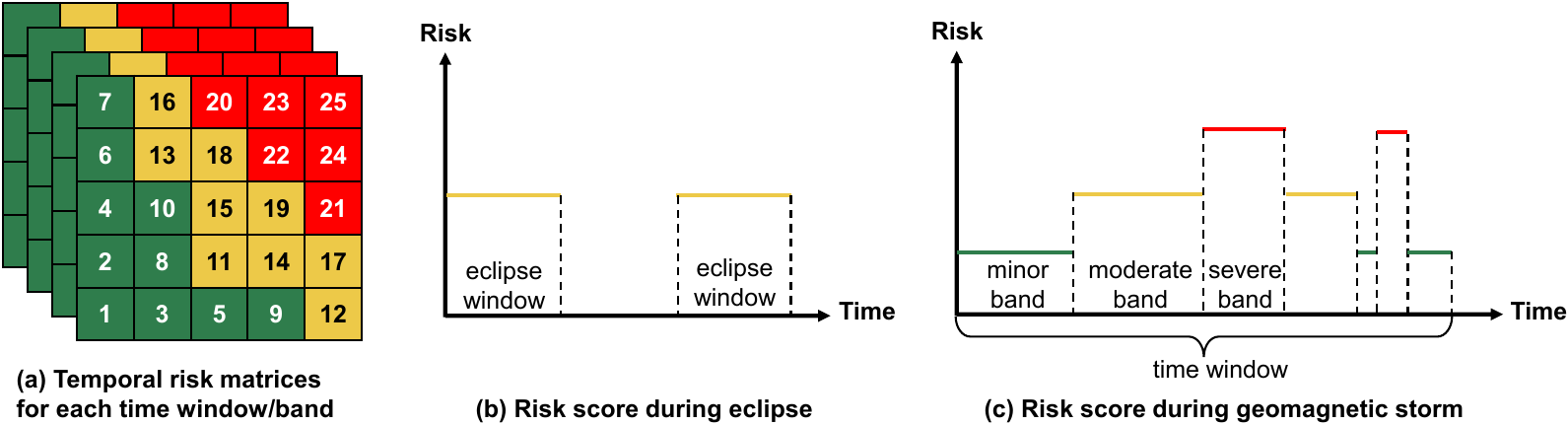}
  \caption{Temporal Risk Matrices and Time-Resolved Risk Across Windows and Bands.}
  \label{fig:window}
\end{figure*}

\noindent \textbf{Scenario-specific timing constraints.}
These constraints arise from the concrete attack objective. They do not change the underlying visibility or mission windows. Instead they restrict how those primitive windows can be used in practice. For example, an adversary may need to align an uplink with a narrow command window or overlap with a critical mission phase such as a maneuver or payload operation. Such constraints shrink and correlate the set of usable windows for that scenario. Intuitively, scenarios with high TED often require a richer temporal capability profile $\langle A, F, S, Y, H\rangle$ on the adversary side. We keep this separation explicit: TED is evaluated from the system and the scenario, while $\langle A, F, S, Y, H\rangle$ is used to determine which attacker tiers can satisfy those timing requirements.

Combining motivation, temporal capabilities, and time-varying exploitation difficulty yields the likelihood score. As in NRS~\cite{ear2024towards}, this likelihood is a semi-quantitative, expert-elicited rating rather than a closed-form expression. The difference is that elicitation now explicitly accounts for temporal considerations.

\subsection{Temporal Risk Matrix}

In real-world systems, both likelihood and impact are time-dependent. Time should not be viewed merely as an external parameter that scales a fixed risk value computed under the assumption of time-invariant likelihood and impact. To stay compatible with the semi-quantitative $5\times5$ risk matrix, we avoid a full three-dimensional risk surface. Instead, at the modeling level we treat likelihood and impact as functions of time, $L(t)$ and $I(t)$. We then project this time-resolved analysis back onto a two-dimensional grid by assigning discrete scores from $1$ to $5$ across the mission timeline. The resulting temporal risk matrix is best viewed as a family of $5\times5$ matrices indexed by time, as shown in \autoref{fig:window}(a), rather than as a single time-agnostic picture of risk.

To keep the number of risk-matrix snapshots tractable, we discretize the mission into time windows, such as passes, mission phases, or other critical intervals. Beyond recurring orbital events, distinct lifecycle phases act as unique risk windows: the Launch and Early Orbit Phase (LEOP) heightens TT\&C vulnerability during orbit raising and stabilization, while Decommissioning reduces vigilance and relaxes security, changing the system's defenses. In the simplest cases, for example eclipse versus non-eclipse as shown in \autoref{fig:window}(b), the relevant time windows are naturally discrete and internally homogeneous. A single $(L,I)$ level can then be assigned to each window. For environmental episodes such as geomagnetic storms or South Atlantic Anomaly transits, as shown in \autoref{fig:window}(c), risk can vary within the episode as a function of storm intensity or radiation flux. In our framework, we further discretize the underlying intensity, for example storm level or particle flux, into a small number of bands and treat each band as a separate time window with its own $(L,I)$ level. Each time window or time-severity band still maps to a well-defined cell in the $5\times5$ risk matrix, while the underlying analysis remains time-resolved and environment-aware.

Likelihood varies across time windows and bands through the combined effect of motivation, static exploitation difficulty, TED, and the minimal attacker tier. Once the mission has been discretized, we assign an ordinal likelihood level from $1$ to $5$ to each window or band. For a fixed technique and attack objective, we treat likelihood as piecewise constant within homogeneous window classes. Windows of the same type, such as routine passes with similar geometry or eclipse intervals with similar temporal characteristics, share the same likelihood level. Likelihood changes when the system moves between window classes, for example from routine operations to eclipse-critical or high storm conditions, or when the attack objective itself changes. Impact varies across windows and bands; it reflects the consequences if the attack succeeds in a given window.

\section{Case Study}

We apply our temporal risk framework to STARMELT~\cite{zhang2023energy} as a concrete case study. Under the SPARTA framework~\cite{ear2024towards}, STARMELT maps to \emph{EX-0013.01: Valid Commands}, where an attacker issues legitimate (protocol-valid) actions in a way that progressively depletes onboard resources such as stored battery energy.

For high-criticality satellite systems, SPARTA yields a static NRS of 25 (High) by taking worst-case Likelihood and Impact. In LEO operations, however, the same technique can have very different consequences depending on the power regime: during sunlight the platform is typically power-positive, whereas during eclipse it operates on battery discharge. Treating the entire orbit as uniformly “High” encourages always-on throttling even when margins are available, while a single “average” score can understate the short eclipse interval where resource depletion is most consequential. To make this dependence explicit without leaving the $5\times5$ matrix, we score Sunlight and Eclipse as separate windows.

\subsection{Likelihood Analysis}
We assess the likelihood of the STARMELT attack based on Temporal Adversary Capabilities and TED. Internet Constellations (SICs) are critical infrastructure, so we assume the adversary's motivation remains high.

\noindent \textbf{Temporal Adversary Capabilities.}
To execute STARMELT, an attacker must possess a specific Minimal Temporal Capability Set ($M_{T}$). We assume an adversary with at least Tier-4 temporal capabilities (per \autoref{tab:capability}), as executing STARMELT requires the following capabilities:

\begin{itemize}
    \item STARMELT relies on predicting when the victim satellite enters and exits eclipse and on understanding constellation topology/routing dynamics. This requires capability $F$ (e.g., ephemeris propagation and lighting prediction) and capability $S$ (sensing system connectivity and exploiting exposed topology/routing logic).

    \item STARMELT employs ``Victim Stalking'' and ``Victim Crossing'' strategies, coordinating geo-distributed bots to inject traffic aligned with fast orbital motion and window transitions. This requires capability $Y$ to align ground asset activation with satellite phase timing.

    \item STARMELT also requires capability $A$. The attacker must be able to generate protocol-conformant traffic and inject it over the victim’s relevant links at a high enough rate. The goal is to sustain congestion
and keep power draw elevated.
\end{itemize}

The required capability set is thus $ M_T = \{A, F, S, Y\}$. For this case study, we treat $M_T$ as constant across sunlight and eclipse windows, since generating protocol-conformant traffic and coordinating timing do not directly depend on illumination. Any illumination-driven operational changes (e.g., routing policy changes) are instead reflected through TED.

\noindent \textbf{Temporal Exploitation Difficulty.}
We distinguish scenario-specific constraints from intrinsic window properties.

\paragraph{Intrinsic properties}
The eclipse window is shorter than the sunlight window, which reduces temporal slack. While the attack vector (protocol-conformant traffic injection) can be initiated on demand and does not require a dedicated warm-up phase, effective execution still depends on reaching a sufficient injection rate and maintaining it long enough within the window. A shorter window therefore provides fewer retries and less margin for timing error, moderately increasing TED compared to sunlight.

\paragraph{Scenario-Specific Constraints}
The primary constraints for STARMELT depend on the attack vector. Ground-to-satellite link (GSL) saturation is geometry-limited, requiring bots to remain within the victim footprint to maintain LoS, while inter-satellite link (ISL) saturation is topology-limited, requiring traffic to traverse feasible routing paths as connectivity evolves. These constraints do not directly depend on illumination, but illumination-driven operational behavior can further shrink the effective usable portion of each window; we account for such effects as scenario-specific constraints when evaluating TED.

Because TED is moderately higher during eclipse due to reduced slack and a potentially smaller effective window, we rate the likelihood as Very High (5) in sunlight and High (4) in eclipse.

\subsection{Impact Analysis}
Unlike Likelihood, the operational Impact is heavily modulated by the satellite’s power state. We align our scoring with standard NASA risk consequence criteria~\cite{nasa2011risk}.

\noindent \textbf{Sunlight Window.} 
During sunlight, the solar arrays generally cover the bus load and allow the battery to recharge. In STARMELT’s illuminated-case evaluation~\cite{zhang2023energy}, sustained traffic flooding did not increase battery depth-of-discharge (DoD), consistent with a net-positive energy balance. Under illumination, the consequences are primarily non-energy, driven by higher onboard computation. This fits the \emph{Minor Degradation} category: performance degrades but stays within system margins. We therefore score Impact as 2 (Low).

\noindent \textbf{Eclipse Window.}
The satellite relies entirely on battery discharge. STARMELT shows that keeping the satellite out of low-power sleep during eclipse increases discharge and accelerates battery wear, reducing effective battery life (over repeated eclipse intervals, up to $\sim$76\% over 1.5 years in the evaluated setting)~\cite{zhang2023energy}. With tight power margins, this may trigger undervoltage protection and safe modes; in the worst case, it can disrupt attitude control and communications. We therefore treat eclipse consequences as \emph{Mission-Threatening} and score Impact = 5 (Very High).

\subsection{Operational Value}

\begin{table}[h]
\centering
\caption{Temporal Risk Comparison (High Criticality System)}
\label{tab:temporal_risk}
\renewcommand{\arraystretch}{1.2}
\begin{tabular}{|l|c|c|c|}
\hline
\textbf{Assessment Mode} & \textbf{Likelihood} & \textbf{Impact} & \textbf{Risk Score} \\ \hline
Static Baseline (SPARTA) & 5 & 5 & \textbf{25 (High)} \\ \hline
\textbf{Temporal (Sunlight)} & 5 & 2 & \textbf{16 (Medium)} \\ \hline
\textbf{Temporal (Eclipse)} & 4 & 5 & \textbf{24 (High)} \\ \hline
\end{tabular}
\end{table}

\autoref{tab:temporal_risk} highlights the temporal split: eclipse carries the highest risk, whereas sunlight is materially lower. In practice, this supports stricter controls during eclipse and less restrictive non-critical throttling during sunlight to preserve throughput, without relaxing baseline safeguards.

\section{Discussion}


\noindent \textbf{Validation and calibration.}
Our framework remains semi-quantitative. Time-resolved likelihood and impact levels are still elicited from experts. An open problem is how to calibrate and validate such temporal risk models. Public space cyber incident data are scarce and incomplete. They rarely reveal how outcomes depend on timing. The community still lacks standardized time-aware red-team exercises, simulations, and benchmarks. These artifacts would help test whether temporal risk profiles produced by our approach align with periods when systems would be most vulnerable.

\noindent \textbf{Temporal attack-defense games.}
We currently treat temporal risk as a property of a given system and adversary, without modeling strategic interaction between attackers and defenders who both choose when to act. An open problem is to embed temporal risk assessment into explicit attack-defense games over time, where attackers select windows to exploit and defenders decide when to harden, monitor, or reschedule operations.

\noindent \textbf{Scalable compositional modeling.}
Our temporal abstractions are currently defined for a single spacecraft and its environment. Future missions increasingly involve constellations, shared ground infrastructure, and multiple services. Enumerating all combinations of time windows, environmental bands, assets, and services does not scale.
However, this complexity also offers an opportunity for defense through \textit{temporal diversity}: since satellites in different orbital planes enter high-risk windows (e.g., SAA transit) at staggered times, future frameworks could enable \textit{risk-aware routing} that dynamically directs traffic away from vulnerable assets.
To achieve this, an open problem is how to aggregate risk at the orbital plane level to build constellation-wide assessments without losing necessary temporal structure.

\section{Conclusion}

In this paper, we argue that security risk for space systems cannot be treated as a static property of individual techniques. It must be evaluated in light of when attacks are feasible and how operations and environment change over time. We complement existing frameworks with a time-sensitive view that augments attacker tiers with a temporal mastery profile and separates static from timing-related difficulty. We then represent risk as a family of likelihood–impact matrices parameterized by operational windows and environmental bands. These elements let operators tie risk assessment more directly to orbital dynamics, mission timelines, and space weather.


\section*{Acknowledgment}

This work was supported by ONR grant N00014-23-1-2122.



%
\bibliographystyle{IEEEtran}
\bibliography{reference}

@misc{starlink2025official,
  author       = {{SpaceX}},
  title        = {{Starlink}: Reliable High-Speed Internet from Space},
  year         = {2025},
  howpublished = {\url{https://www.starlink.com/}},
  note         = {Accessed: 2025-11-16}
}

@misc{aws2025groundstation,
  author       = {{Amazon Web Services}},
  title        = {{AWS Ground Station}},
  year         = {2018},
  howpublished = {\url{https://aws.amazon.com/ground-station/}},
  note         = {Accessed: 2025-11-16}
}

@misc{azure2025orbital,
  author       = {{Microsoft Azure}},
  title        = {{Azure Orbital Ground Station}},
  year         = {2022},
  howpublished = {\url{https://azure.microsoft.com/en-us/products/orbital}},
  note         = {Accessed: 2025-11-16}
}

@inproceedings{willbold2023space,
  author    = {Willbold, Johannes and Schloegel, Moritz and V{\"o}gele, Manuel and Gerhardt, Maximilian and Holz, Thorsten and Abbasi, Ali},
  title     = {Space Odyssey: An Experimental Software Security Analysis of Satellites},
  booktitle = {2023 {IEEE} Symposium on Security and Privacy ({SP})},
  pages     = {1--19},
  year      = {2023},
  publisher = {{IEEE}}
}

@misc{viasat2022attack,
  author       = {{Viasat Inc.}},
  title        = {{KA-SAT} Network Cyber Attack Overview},
  year         = {2022},
  month        = mar,
  howpublished = {Viasat Corporate Blog},
  note         = {\url{https://www.viasat.com/perspectives/corporate/2022/ka-sat-network-cyber-attack-overview/} (accessed 2025-11-16)}
}

@misc{mitre_attack_website,
  author       = {{The MITRE Corporation}},
  title        = {{MITRE ATT\&CK\textregistered}},
  year         = {2025},
  howpublished = {\url{https://attack.mitre.org/}},
  note         = {Accessed: 2025-11-16}
}

@misc{rhodes2019skyfield,
  author       = {Rhodes, Brandon},
  title        = {Skyfield: High Precision Research-Grade Positions for Planets and Earth Satellites Generator},
  year         = {2019},
  howpublished = {\url{https://ascl.net/1907.024}},
  note         = {Astrophysics Source Code Library, record ascl:1907.024}
}

@techreport{sawyer1976ap,
  author      = {Sawyer, Donald M. and Vette, James I.},
  title       = {AP-8 Trapped Proton Environment for Solar Maximum and Solar Minimum},
  institution = {NASA Goddard Space Flight Center},
  year        = {1976},
  number      = {NSSDC/WDC-A-R\&S 76-06}
}

@article{peled2023evaluating,
  author  = {Peled, Roy and Aizikovich, Eran and Habler, Edan and Elovici, Yuval and Shabtai, Asaf},
  title   = {Evaluating the Security of Satellite Systems},
  journal = {arXiv preprint arXiv:2312.01330},
  year    = {2023}
}

@inproceedings{remy2025sok,
  author    = {Remy, Jose Luis Castanon and Ear, Ekzhin and Chang, Caleb and Feffer, Antonia and Xu, Shouhuai},
  title     = {So{K}: Space Infrastructures Vulnerabilities, Attacks and Defenses},
  booktitle = {2025 {IEEE} Symposium on Security and Privacy ({SP})},
  pages     = {1028--1046},
  year      = {2025},
  publisher = {{IEEE}}
}

@misc{aerospace_sparta_website,
  author       = {{The Aerospace Corporation}},
  title        = {{Space Attack Research \& Tactic Analysis (SPARTA)}},
  year         = {2025},
  howpublished = {\url{https://sparta.aerospace.org/}},
  note         = {Accessed: 2025-11-16}
}

@inproceedings{willbold2025space,
  author    = {Willbold, Johannes and Cloosters, Tobias and W{\"o}rner, Simon and Buchmann, Felix and Schloegel, Moritz and Davi, Lucas and Holz, Thorsten},
  title     = {Space {RADSIM}: Binary-Agnostic Fault Injection to Evaluate Cosmic Radiation Impact on Exploit Mitigation Techniques in Space},
  booktitle = {2025 {IEEE} Symposium on Security and Privacy ({SP})},
  pages     = {1047--1063},
  year      = {2025},
  publisher = {{IEEE}}
}

@inproceedings{tibaldo2025gnss,
  author    = {Tibaldo, Christopher and Sathaye, Harshad and Camurati, Giovanni and Capkun, Srdjan},
  title     = {{GNSS-WASP}: {GNSS} Wide Area {SP}oofing},
  booktitle = {{USENIX} Security 2025},
  pages     = {7039--7058},
  year      = {2025}
}

@inproceedings{salkield2023satellite,
  author    = {Salkield, Edd and Szak{\'a}ly, Marcell and Smailes, Joshua and K{\"o}hler, Sebastian and Birnbach, Simon and Strohmeier, Martin and Martinovic, Ivan},
  title     = {Satellite Spoofing from {A} to {Z}: On the Requirements of Satellite Downlink Overshadowing Attacks},
  booktitle = {Proceedings of the 16th {ACM} Conference on Security and Privacy in Wireless and Mobile Networks ({WiSec} '23)},
  pages     = {341--352},
  year      = {2023},
  publisher = {{ACM}}
}

@inproceedings{oligeri2020gnss,
  author    = {Oligeri, Gabriele and Sciancalepore, Savio and Di Pietro, Roberto},
  title     = {{GNSS} Spoofing Detection via Opportunistic {IRIDIUM} Signals},
  booktitle = {Proceedings of the 13th {ACM} Conference on Security and Privacy in Wireless and Mobile Networks ({WiSec} '20)},
  pages     = {42--52},
  year      = {2020},
  publisher = {{ACM}}
}

@inproceedings{wouters2022glitched,
  author    = {Wouters, Lennert},
  title     = {Glitched on Earth by Humans: A Black-Box Security Evaluation of the {SpaceX} Starlink User Terminal},
  booktitle = {Black Hat USA 2022},
  year      = {2022},
  note      = {Conference talk and whitepaper}
}

@inproceedings{willbold2024vsaster,
  author    = {Willbold, Johannes and Schloegel, Moritz and Bisping, Robin and Strohmeier, Martin and Holz, Thorsten and Lenders, Vincent},
  title     = {{VSAsTer}: Uncovering Inherent Security Issues in Current {VSAT} System Practices},
  booktitle = {Proceedings of the 17th {ACM} Conference on Security and Privacy in Wireless and Mobile Networks ({WiSec} '24)},
  pages     = {288--299},
  year      = {2024},
  publisher = {{ACM}},
  doi       = {10.1145/3643833.3656139}
}

@article{smailes2023dishing,
  author  = {Smailes, Joshua and Salkield, Edd and K{\"o}hler, Sebastian and Birnbach, Simon and Martinovic, Ivan},
  title   = {Dishing out {DoS}: How to Disable and Secure the Starlink User Terminal},
  journal = {arXiv preprint arXiv:2303.00582},
  year    = {2023}
}

@inproceedings{scharnowski2023case,
  author       = {Scharnowski, Tobias and Buchmann, Felix and W{\"o}rner, Simon and Holz, Thorsten},
  title        = {A Case Study on Fuzzing Satellite Firmware},
  booktitle    = {Proceedings of the 2023 Workshop on the Security of Space and Satellite Systems ({SpaceSec}'23)},
  year         = {2023},
  address      = {San Diego, CA, USA},
  publisher    = {Internet Society},
  doi          = {10.14722/spacesec.2023.230707},
  url          = {https://www.ndss-symposium.org/wp-content/uploads/2023/06/spacesec2023-230707-paper.pdf}
}

@article{vanlyssel2025spychain,
  author  = {Vanlyssel, Jack and Sobrados, Enrique and Anwar, Ramsha and Roman, Gruia{-}Catalin and Anwar, Afsah},
  title   = {SpyChain: Multi-Vector Supply Chain Attacks on Small Satellite Systems},
  journal = {arXiv preprint arXiv:2510.06535},
  year    = {2025}
}

@article{marra2024feasibility,
  author  = {Marra, Gabriele and Planta, Ulysse and W{\"u}stenberg, Philipp and Abbasi, Ali},
  title   = {On the Feasibility of CubeSats Application Sandboxing for Space Missions},
  journal = {arXiv preprint arXiv:2404.04127},
  year    = {2024}
}

@inproceedings{young2010initial,
  author       = {Young, Anthony and Kitts, Christopher and Neumann, Michael and Mas, Ignacio and Rasay, Mike},
  title        = {Initial Flight Results for an Automated Satellite Beacon Health Monitoring Network},
  booktitle    = {Proceedings of the 24th Annual {AIAA}/{USU} Conference on Small Satellites},
  year         = {2010},
  address      = {Logan, UT, USA},
  organization = {{AIAA}/{USU}},
  note         = {SSC10-XII-1}
}

@inproceedings{yadav2024orbital,
  author       = {Yadav, Nikita and Vollmer, Franziska and Sadeghi, Ahmad{-}Reza and Smaragdakis, Georgios and Voulimeneas, Alexios},
  title        = {Orbital Shield: Rethinking Satellite Security in the Commercial Off-the-Shelf Era},
  booktitle    = {2024 Security for Space Systems ({3S})},
  pages        = {59--69},
  year         = {2024},
  organization = {{IEEE}}
}

@inproceedings{wyatt1997overview,
  author    = {Wyatt, E. Jay and Foster, Mike and Schlutsmeyer, Alan and Sherwood, Rob and Sue, Miles K.},
  title     = {An Overview of the Beacon Monitor Operations Technology},
  booktitle = {International Symposium on Artificial Intelligence, Robotics, and Automation in Space},
  year      = {1997}
}

@article{ear2024towards,
  author  = {Ear, Ekzhin and Bailey, Brandon and Xu, Shouhuai},
  title   = {Towards Principled Risk Scores for Space Cyber Risk Management},
  journal = {arXiv preprint arXiv:2402.02635},
  year    = {2024}
}

@inproceedings{kurii2022analysis,
  author    = {Kurii, Yevhenii and Opirskyy, Ivan},
  title     = {Analysis and Comparison of the {NIST} {SP} 800-53 and {ISO}/{IEC} 27001:2013},
  booktitle = {Proceedings of the 2nd International Workshop on Cyber Protection of Information and Telecommunication Systems ({CPITS} 2022)},
  year      = {2022},
  pages     = {21--32},
  series    = {{CEUR} Workshop Proceedings},
  volume    = {3288},
  publisher = {{CEUR}-{WS}.org}
}

@techreport{bailey2021cybersecurity,
  author      = {Bailey, Brandon},
  title       = {Cybersecurity Protections for Spacecraft: A Threat Based Approach},
  institution = {The Aerospace Corporation},
  year        = {2021},
  number      = {TOR-2021-01333-Rev A},
  month       = apr,
  note        = {\url{https://aerospace.org/paper/cybersecurity-protections-spacecraft-threat-based-approach}}
}

@inproceedings{zhang2023energy,
  author       = {Zhang, Yaoying and Wu, Qian and Lai, Zeqi and Deng, Yangtao and Li, Hewu and Li, Yuanjie and Liu, Jun},
  title        = {Energy Drain Attack in Satellite Internet Constellations},
  booktitle    = {2023 {IEEE}/{ACM} 31st International Symposium on Quality of Service ({IWQoS})},
  pages        = {1--10},
  year         = {2023},
  organization = {{IEEE}}
}

@techreport{nasa2011risk,
  author       = {{National Aeronautics and Space Administration}},
  title        = {NASA Risk Management Handbook},
  institution  = {National Aeronautics and Space Administration},
  year         = {2011},
  number       = {NASA/SP-2011-3422},
  month        = nov,
  note         = {Version 1.0. \url{https://www.nasa.gov/wp-content/uploads/2023/08/nasa-risk-mgmt-handbook.pdf}}
}

\end{document}